# Test results of the LQXFA/B02 and LQXFA/B03 cryo-assemblies for the High Luminosity LHC upgrade

G. Chlachidze, G. Ambrosio, M. Anerella, G. Apollinari, M. Baldini, J. DiMarco, S. Feher, P. Ferracin,
V. Nikolic, F. Nobrega, C. Orozco, D. Orris, S. Prestemon, R. Rabehl, S. Stoynev, A. Vouris

*Abstract*—**The US High-Luminosity LHC Accelerator Upgrade Project (AUP) is responsible for delivering cryo-assemblies for the Q1/Q3 quadrupole optical components of the High Luminosity LHC upgrade at CERN. Total of 10 cryo-assemblies containing two $Nb_3Sn$ quadrupole magnets per cold mass will be delivered within this program. After the successful test of the first pre-series cryo-assembly in 2023, two more cryo-assemblies were tested at Fermilab's horizontal test facility.**

**Production overview and the test results of the LQXFA/B02 and LQXFA/B03 cryo-assemblies are summarized in this paper. After the first test, to increase the capability of the horizontal test facility, various improvements have been made. These improvements are also described in this paper.**

## I. INTRODUCTION

THE US High-Luminosity LHC Accelerator Upgrade Project (AUP) is delivering a total of 10 cryo-assemblies with the Q1/Q3 quadrupole components for the High-Luminosity LHC upgrade at CERN [1-2]. Two 4.2-m long and 150-mm aperture $Nb_3Sn$ quadrupole magnets [3] are used in each cryo-assembly. Before the magnets are integrated into a helium vessel (cold mass) and inserted in a vacuum vessel, they are all tested vertically at Brookhaven National Laboratory (BNL). Details on cold mass and cryo-assembly fabrication at Fermilab can be found in [3-4]. Then these cryo-assemblies are horizontally tested at Fermilab's magnet test facility [5].

The first pre-series cryo-assembly LQXFA/B01 was tested at Fermilab in 2023, and test results were presented in [6]. Two more cryo-assemblies have been tested since then – LQXFA/B02 in 2024 and LQXFA/B03 in 2025. The test

Submitted for review July 28, 2025. This work was supported by the U.S. Department of Energy, Office of Science, Office of High Energy Physics, through the US LHC Accelerator Upgrade Project (AUP). (*Corresponding author: Guram Chlachidze*, e-mail: guram@fnal.gov).

G. Chlachidze, G. Ambrosio, G. Apollinari, M. Baldini, J. DiMarco, S. Feher, V. Nikolic, F. Nobrega, C. Orozco, D. Orris, R. Rabehl, S. Stoynev and A. Vouris are with Fermi National Accelerator Laboratory, Batavia, IL 60510 USA

M. Anerella is with Brookhaven National Laboratory, Upton, NY 11973-5000 USA

P. Ferracin and S. Prestemon are with Lawrence Berkeley National Laboratory, Berkeley, CA 94720 USA

results of these two cryo-assemblies are summarized in this paper. After the first test, to increase the capability of the horizontal test facility, various improvements were made at Fermilab. The most relevant improvements are also described in this paper. A picture of the LQXFA/B02 installation on the horizontal test stand is shown in Fig. 1.

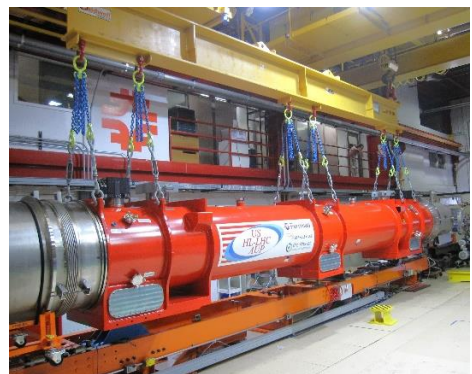

**Fig. 1.** Installation of LQXFA/B02 at Fermilab's horizontal test stand.

## II. HORIZONTAL TEST STAND IMPROVEMENTS

The LQXFA/B01 test was used as a final commissioning of the upgraded horizontal test facility at Fermilab. Some test stand capabilities, including controlled cooldown and warmup, helium recovery after a quench, and operation at high pressures were commissioned with a cryo-assembly for the first time. This test also demonstrated that more improvements were required to achieve the project goals for cryo-assemblies horizontal test. Two major improvements in preparation for LQXFA/B02 test are described below.

The liquefying capacity of the cryogenic plant at the magnet test facility was not sufficient for the continuous operation of the horizontal test stand at 1.9 K. A few years ago, anticipating the increase of the testing activities related to the superconducting magnet and cavity development programs at Fermilab, an upgrade of the existing helium cryogenic plant was initiated. This upgrade included a new cryogenic helium liquefier (cold box) from *Air Liquid Advance Technologies* (ALAT) and a 4000-liter liquid helium storage dewar [7]. The ALAT cold box was commissioned in 2023 and reached its full operation capacity in 2024, at the same time that LQXFA/B02 was delivered for testing. The old CTI-1500 and new ALAT liquefiers combined can provide up to 600 liters of liquid helium per hour, eliminating any issues with liquid helium inventory for the horizontal cryo-assembly testing. The liquid helium storage volume reached ~14,000 liters after the upgrade.

During the powering tests of LQXFA/B01, the dynamic heat load to the 4.5 K feed box was found to be high. Heat load increase of 200 W was measured when the magnets were ramped from 6 kA to 16 kA. The issue was caused by a high resistance lead joint at the bottom flag in one of the power leads (shown in Fig. 3, left). To avoid trips due to overheating the lead joint, we had to increase the liquid helium level in the feed box, significantly increasing consumption of liquid helium.

To repair the faulty lead joint, a new copper fixture encapsulating the bottom flag was designed and fabricated at Fermilab. Pure (oxygen-free) copper was used to fabricate two parts of the fixture (see Fig. 2, left). The longer lead joint assembly would increase the portion of the flag submerged in liquid helium, improving the cooling conditions. To further improve thermal contact, the new fixture was soldered onto the flag. Six 1 kW heater cartridges (Watlow 2348-7390) were used for soldering. Before applying to the test stand, the newly designed lead joint was bench tested using a mockup of the bottom flag. The test assembly was sectioned in a few layers to verify the quality of the soldering (see Fig. 2, right).

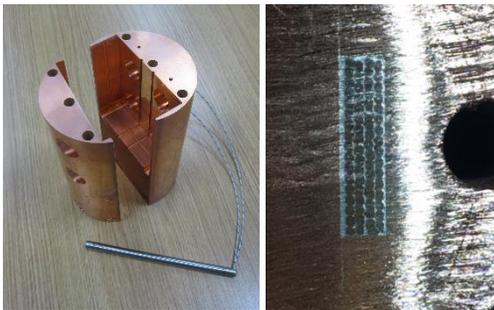

**Fig. 2.** The new copper fixtures and a 1 kW heater cartridge (left) and a cross section of the soldering area (right)

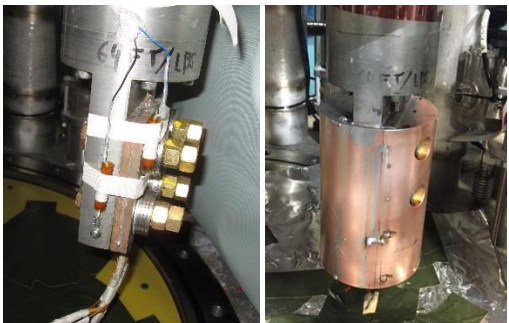

**Fig. 3.** The original (left) and the upgraded (right) lead joint assembly

After the installation of the copper fixture at the stand, dynamic heat load measurements were taken during powering of LQXFA/B02. The dynamic heat load to the feed box was reduced from 200 W to 12.6 W [8]. Fig. 3 shows the original joint assembly (left) and the same joint after the upgrade (right).

## III. Controlled Cooldown and Warmup of LQXFA/B02 and LQXFA/B03

To prevent excessive stresses in the magnet structure during cooldown and warmup, the maximum temperature difference between the ends of each magnet cannot exceed 50 K. For monitoring temperature difference, temperature sensors (RTD) are installed on both ends of each magnet (see Fig. 4).

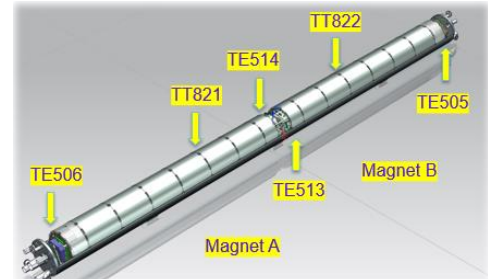

**Fig. 4.** Location of the RTD used for monitoring of the temperature gradient in two magnets of the cryo-assembly

In LQXFA/B03 the middle RTDs TE513 and TE514 were not installed, therefore the temperature gradient (delta T) was calculated using the (TE505, TT822) and (TT821, TE506) pairs. Since these RTD pairs cover only 2/3$^{rd}$ of the magnet length, the temperature gradient limit was reduced from 50 K to 40 K in LQXFA/B03 only.

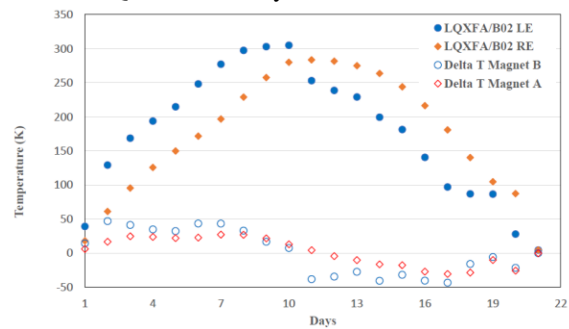

**Fig. 5.** Controlled warmup and cooldown of LQXFA/B02

The temperature change at the cryo-assembly lead and return ends (LE and RE respectively) during the thermal cycle in both cryo-assemblies is shown in Figs. 5 and 6. The temperature difference across of each magnet is also shown. The controlled cooldown or warmup alone takes 9-10 days. To make it shorter, we need to significantly increase the helium gas flow in the helium supply line from currently available 16-18 g/s, which is limited by the existing heat exchanger in the helium return line.

## IV. LQXFA/B02 and LQXFA/B03 Test Results

Both magnets in LQXFA/B02 (MQXFA05 and MQXFA06) and LQXFA/B03 (MQXFA10 and MQXFA11) were previously trained to the acceptance current of 16530 A at BNL [9]. The nominal operating current for these magnets is 16230 A. All qualification tests at Fermilab are performed at nominal operating temperature of 1.9 K. Temperature dependence test was performed at about 4 K, when single-phase helium provides better cooling conditions. Tests were performed in two thermal cycles, with a warmup to room temperature and cooldown to 1.9 K in between.

Magnet protection is based on heater strips and Coupling-Loss Induced Quench (CLIQ) system [10]. The external energy extraction system was removed from the quench protection. Test parameters and procedures are described in [6].

The field quality measurements will be summarized in a separate publication.

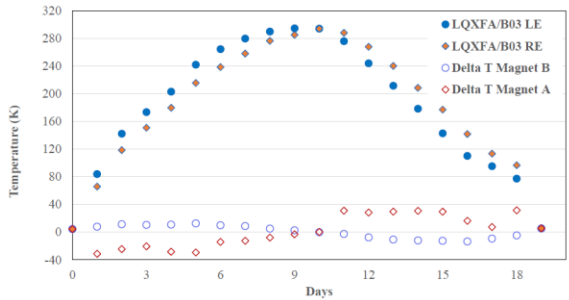

**Fig. 6.** Controlled warmup and cooldown of LQXFA/B03

### A. LQXFA/B02 Quench Performance

Quench performance of LQXFA/B02 is shown in Fig. 7. The acceptance current was reached in one quench, and it was the only quench in two thermal cycles. The quench was detected in MQXFA06 at 16524 A. In the next ramp, we successfully reached the acceptance current of 16530 A with a standard ramp rate of 20 A/s, and stayed at the flattop for 30 min. Following ramp to the nominal current of 16230 A at 30 A/s ramp up rate (5 min at the flat-top) and 150 A/s ramp down rate was successful. Completing the first thermal cycle (TC1), we performed the temperature dependence test at 4.0 K - magnets were successfully ramped to the nominal current and stayed at flattop for 30 min.

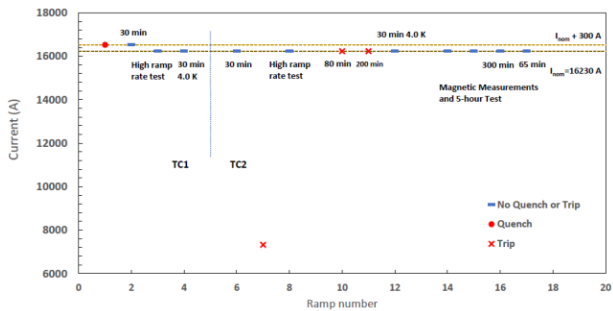

**Fig. 7.** LQXFA/B02 quench performance. All high current ramps are shown, including those for magnetic measurements.

In the second thermal cycle (TC2), LQXFA/B02 reached the nominal current without a quench. During the high ramp test, a few trips occurred due to power supply related spikes in the quench detection signals. No resistive signal was detected in the magnets. By adjusting the quench detection thresholds, the remaining run plan was completed successfully, including the 5-hour holding current test at the nominal current of 16230 A.

### B. LQXFA/B03 Quench Performance

LQXFA/B03 quench performance is shown in Fig. 8. No powering tests were performed in TC1 due to a breakdown of the coil-to-ground electrical insulation during standard high voltage qualification tests at 1.9 K. After extensive investigation, the ground fault was associated with one of the voltage taps, located very close to the instrumentation feedthrough. The cryo-assembly rework was done at the test facility, without disassembly of the connections between the test stand and the cryo-assembly. After the rework, LQXFA/B03 passed all the high-voltage qualification tests according to the electrical design criteria for the HL-LHC inner triplet magnets [11].

In TC2, the acceptance current was reached in the first ramp, but almost immediately the quench detection tripped due to the cooling issue in the water-cooled flexible power leads. It was discovered that the cooling lines were clogged with miniature balls of the deionizing material used in the low conductivity water system. Before the problem was identified and fixed, another trip was detected at about the nominal current. Then the only spontaneous quench occurred in MQXFA10 at 16433 A. All the remaining tests, including the high ramp rate test, 5-hour holding current test at the nominal current, and the temperature dependence test at 4.1 K were successfully completed.

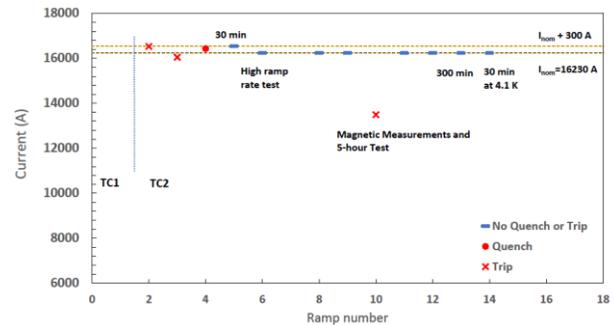

**Fig. 8.** LQXFA/B03 quench performance. All high current ramps are shown, including those for magnetic measurements.

### B. Quench Integral Estimates

Accumulated quench integrals (integral of current squared vs. time from the quench start) during the horizontal test of LQXFA/B02 and LQXFA/B03 are summarized in Fig. 9. Two spontaneous quenches and the trips at 6000 A or higher are included in this plot. The estimated quench integral for these two quenches is less than 27.3 MIITs, i.e. below the quench integral limit of 32 MIITs, corresponding to the maximum allowable hot-spot temperature of 250 K [12].

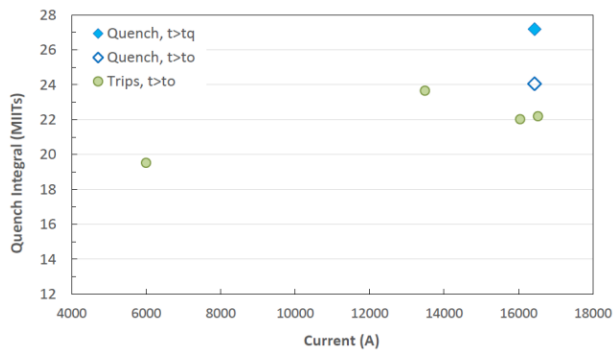

**Fig. 9.** Quench integrals in LQXFA/B02 and LQXFA/B03

### C. Splice resistance measurements

According to the AUP cold mass and cryo-assembly acceptance criteria [13], all superconductor splices must have resistance less than 1 n$\Omega$ at 1.9 K. We measured NbTi-NbTi splices made during the cryo-assembly fabrication at Fermilab, which includes splices in the busbars connecting two magnets to each other and to the test facility power leads.

Splice measurements were performed at currents up to 8000 A (stairsteps when ramping up and down). All measured splices satisfy the acceptance criteria. The splice resistance varies between 0.29-0.45 n$\Omega$, except for one with a resistance of 0.77 n$\Omega$. Splice resistance uncertainties are estimated as 0.06 n$\Omega$.

Examples of the splice resistance measurement in the busbar of LQXFA/B02 and LQXFA/B03 are shown in Fig. 10.

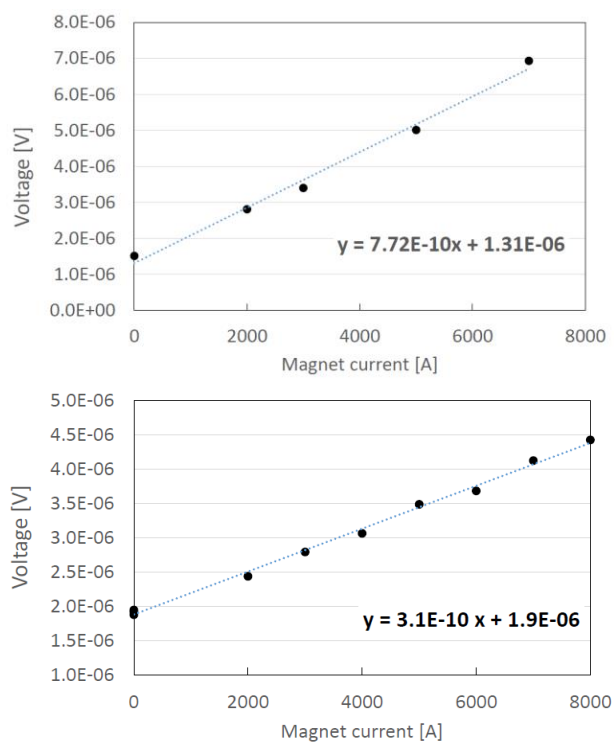

**Fig. 10.** Examples of the busbar splice measurements in LQXFA/B02 (top) and LQXFA/B03 (bottom)

### D. Fringe field measurements

The magnetic fringe field was measured at 10 mm from the cryostat at nominal operating conditions (NOC), i.e. when magnets are at the nominal operating current of 16230 A and at 1.9 K temperature.

The acceptance requirement [14] and measured fringe field values are shown in Table 1.

#### TABLE I
#### MAGNETIC FRINGE FIELD MEASUREMENTS

|  | [Unit] | Required at NOC | LQXFA/B02 | LQXFA/B03 |
|---|---|---|---|---|
| Fringe Field at NOC | mT | < 50 | 2.8 | 3.6 |

### V. CRYO-ASSEMBLY PRODUCTION STATUS

Currently four cryo-assemblies are completed, and three more cryo-assemblies are progressing at the manufacturing site. LQXFA/B05 and LQXFA/B06 cryo-assemblies will be completed in 2025. The cold mass assembly of LQXFA/B07 was started recently.

LQXFA/B01, LQXFA/B02 and LQXFA/B03 have been horizontally tested. The first two cryo-assemblies already have been delivered to CERN and will be included in the HL-LHC inner triplet string test [15].

### V. CONCLUSION

LQXFA/B02 and LQXFA/B03 cryo-assemblies were tested successfully at Fermilab's horizontal magnet test facility. Quench performance of both cryo-assemblies meets the HL-LHC acceptance criteria. Magnets reached the acceptance current of 16530 A in only one quench and demonstrated excellent quench memory after a thermal cycle. All the qualification tests at the nominal operating conditions were passed successfully, including the high ramp rate test and 5-hour holding current test. Both cryo-assemblies passed the temperature dependence test and reached the nominal current at ~ 4 K without a quench.

New liquefier at Fermilab's magnet test facility significantly increased the liquid helium production and storage capabilities, allowing for uninterrupted operation of the horizontal test stand at 1.9 K. The lead joint improvement for one of the power leads significantly reduced the dynamic heat load during high current ramps.

FERMILAB-CONF-25-0526-TD